\documentclass[conference]{IEEEtran}

\usepackage{cite}
\usepackage{amsmath,amssymb,amsfonts}
\usepackage{algorithmic}
\usepackage{graphicx}
\usepackage{textcomp}
\usepackage{xcolor}
\def\BibTeX{{\rm B\kern-.05em{\sc i\kern-.025em b}\kern-.08em
    T\kern-.1667em\lower.7ex\hbox{E}\kern-.125emX}}

\usepackage{xspace}
\usepackage{flushend}
\usepackage[normalem]{ulem}
\usepackage{tikz}

\usepackage[draft]{changes} 

\long\def\ignore#1{}
\sloppypar

\newcommand*\circled[1]{\tikz[baseline=(char.base)]{
  \node[shape=circle,draw,fill=black,text=white,font=\bf,inner sep=0.5pt] (char)
  {\scriptsize#1};
}}

\newcommand{\ApproxSign}{\raise.17ex\hbox{$\scriptstyle\sim$}}

\newcommand{\putsec}[2]{\vspace{-0.0in}\section{#2}\label{sec:#1}\vspace{-0.0in}}
\newcommand{\putssec}[2]{\vspace{-0.0in}\subsection{#2}\label{ssec:#1}\vspace{-0.0in}}
\newcommand{\putsssec}[2]{\vspace{-0.0in}\subsubsection{#2}\label{sssec:#1}\vspace{-0.0in}}

\newcommand{\tabref}[1]{Table~\ref{#1}}
\newcommand{\figref}[1]{Figure~\ref{#1}}
\newcommand{\secref}[1]{Section~\ref{sec:#1}}
\newcommand{\ssecref}[1]{Section~\ref{ssec:#1}}
\newcommand{\sssecref}[1]{Section~\ref{sssec:#1}}

\newcommand{\squishlist}{
   \begin{list}{$\bullet$}
    { \setlength{\itemsep}{0pt}      \setlength{\parsep}{0pt}
      \setlength{\topsep}{3pt}       \setlength{\partopsep}{0pt}
      \setlength{\listparindent}{-2pt}
      \setlength{\itemindent}{-5pt}
      \setlength{\leftmargin}{1em} \setlength{\labelwidth}{0em}
      \setlength{\labelsep}{0.5em} } }

\newcommand{\squishend}{
    \end{list}  }

\newcommand{\squishlisttwo}{
   \begin{list}{$\bullet$}
    { \setlength{\itemsep}{0pt}    \setlength{\parsep}{0pt}
      \setlength{\topsep}{0pt}     \setlength{\partopsep}{0pt}
      \setlength{\leftmargin}{2em} \setlength{\labelwidth}{1.5em}
      \setlength{\labelsep}{0.5em} } }

\newcommand{\PNAME}{\mbox{C3}\xspace}
\newcommand{\PNAMEBOLD}{\mbox{\textbf{C3}}\xspace}
\newcommand{\OPNAME}{\mbox{ConCCL}\xspace}
\newcommand{\OPNAMEBOLD}{\mbox{\textbf{ConCCL}}\xspace}

\usepackage{soul}
\usepackage{letltxmacro}
\LetLtxMacro{\oldhl}{\hl}
\renewcommand{\hl}[1]{\oldhl{#1}}               %Turn on highlighting when review needed

\begin{document}

\title{Optimizing ML Concurrent Computation and Communication with GPU DMA Engines}

\author{

    \IEEEauthorblockN{Anirudha Agrawal}
    \IEEEauthorblockA{
    \textit{Advanced Micro Devices, Inc.}\\
    anirudha.agrawal@amd.com}
    
    \and
    
    \IEEEauthorblockN{Shaizeen Aga}
    \IEEEauthorblockA{
    \textit{Advanced Micro Devices, Inc.}\\
    shaizeen.aga@amd.com}
    
    \and
    
    \IEEEauthorblockN{Suchita Pati}
    \IEEEauthorblockA{
    \textit{Advanced Micro Devices, Inc.}\\
    suchita.pati@amd.com}
    
    \and
    
    \IEEEauthorblockN{Mahzabeen Islam}
    \IEEEauthorblockA{
    \textit{Advanced Micro Devices, Inc.}\\
    mahzabeen.islam@amd.com}
}

\maketitle

\begin{abstract}
Concurrent computation and communication (\PNAME) is a pervasive paradigm in ML and other domains, making its performance optimization crucial. In this paper, we carefully characterize \PNAME in ML on GPUs, which are most widely deployed for ML training and inference. We observe that while \PNAME leads to performance uplifts, the uplifts are far lower than ideal speedups (serial computation and communication versus maximum of computation or communication; all times from isolated executions). That is, \PNAME on average achieves only 21\% of ideal speedup. This is so, due to known challenges of compute and memory interference between concurrent GPU kernels (that is, sharing of GPU's compute units, caches and HBM).

To attain better performance for \PNAME, first, we evaluate dual strategies of schedule prioritization and careful resource partitioning of compute units on GPUs to push performance attained with \PNAME (on average 42\% of ideal speedup). We also provide heuristics that can guide a runtime while employing these strategies. To further enhance \PNAME performance, we propose to mitigate \PNAME interference by offloading communication tasks to the GPU's DMA engines. To this end, we build concurrent communication collectives (\OPNAMEBOLD) proof-of-concepts that harness DMA engines for communication. We show how \OPNAME considerably closes the gap between realized and ideal speedup for \PNAME (on average 72\% of ideal speedup is realized, up to 1.67$\times$ speedup). Overall, our work makes a strong case for GPU DMA engine advancements to better support \PNAME on GPUs. 

\end{abstract}


%%%%%% -- PAPER CONTENT STARTS-- %%%%%%%%
\putsec{sec}{Introduction}
\label{sec:intro}

Large-scale machine learning (ML) models continue to harness distributed computing over increasingly large clusters of GPUs. A consequence of this is intermingling of computation and communication which, in absence of data dependencies, can be executed concurrently on the GPU. There are several ML algorithmic choices that lead to concurrent computation and communication, termed \PNAMEBOLD in this work, such as data-parallelism~\cite{dataParallel90}, fully sharded data parallel (FSDP)~\cite{zhao2023fsdp}, nano-batching~\cite{zhu2024nanoflow} and more. As such, characterization of and optimization for \PNAME on GPUs is important.

To achieve this, we first create a taxonomy of \PNAME and provide a detailed characterization of \PNAME anchored on this taxonomy. Given our ML focus, we study \PNAME manifestations wherein the computation kernel is a matrix-matrix multiplication (or GEMM) kernel and the communication kernel is a collective operation such as an all-gather among multiple GPUs. By analyzing C3 scenarios from the training of the LLaMA-70B and LLaMA-405B~\cite{llama3-24} models, along with some synthetic \PNAME scenarios, we provide broad coverage for our proposed taxonomy.

\begin{figure}[t]
    \centering
    \includegraphics[width=\linewidth]{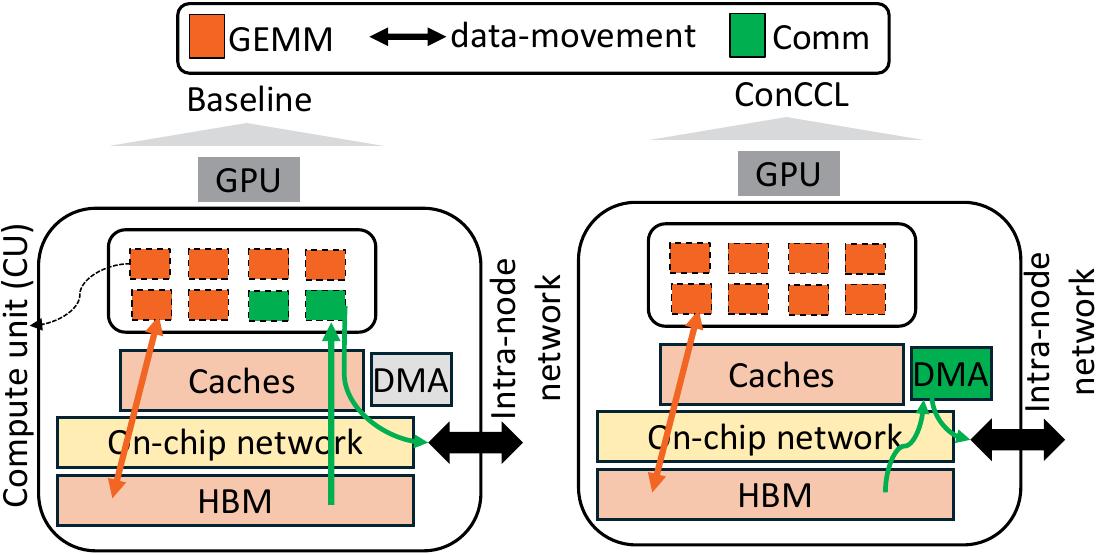}
    \caption{Baseline \PNAME (left) and \PNAME with \OPNAME via DMA offloads (right).}
    \label{fig:c3_overview}
\end{figure}

Next, using detailed experiments, we determine the compute and memory bandwidth requirements of isolated executions of the computation and communication kernels under study. This enables us to get an assessment of ideal performance uplift via concurrency. We observe that, when it comes to compute needs, on a state-of-the-art GPU such as AMD Instinct\textsuperscript{\texttrademark} MI300X which has ample compute units (or GPU compute cores), communication kernels need only up to 10-20\% of available units, making the rest available to concurrent computation kernels (\figref{fig:c3_overview}, left). GEMMs, as expected, manifest a spectrum of behaviors, some resilient to losing compute units to concurrent communication while the rest are sensitive. On the memory front, except for memory bandwidth-bound GEMM kernels, we observe that the bandwidth needs of compute-bound GEMMs and communication kernels can be met in tandem by the high memory bandwidth made available by MI300X GPUs. Overall, for \PNAME scenarios under study, compared to a baseline which serializes computation and communication, if the shorter of the computation or communication kernels is completely hidden, ideal speedups of 1.6$\times$ (average) and 2$\times$ (maximum) are possible.

However, here we observe that of this ideal speedup, on average only 21\% (1.13$\times$) is realized. This is not unexpected, as concurrent GPU kernels cause mutual interference with each other as they share compute cores, caches, high bandwidth memory (HBM) bandwidth and more. That is, the available resources for a given GPU kernel are lower in the presence of other kernels than in isolation. In fact, prior work~\cite{hwang2023ark} has observed that due to such interference, \PNAME can even lead to slowdowns and loss of ML throughput.

To bridge the above performance gap, we first evaluate two strategies: schedule prioritization and careful partitioning of GPU compute units. Specifically, using our isolated execution analysis above, we observe that prioritizing scheduling of kernels with lower resource requirements prevents starvation and leads to overall better performance. Similarly, we also demonstrate that the resource partitioning available on MI300X GPUs, wherein certain compute units on GPUs can be exclusively marked for a given stream of work, can be harnessed to push \PNAME performance. Overall, using these strategies, we can push the performance achieved with \PNAME to an average of about 42\% of available ideal speedup. Additionally, we also provide heuristics that can guide a runtime in employing these strategies.

Finally, to further enhance \PNAME performance, we tackle the interference incurred with \PNAME manifestations today. That is, for compute interference, instead of splitting available compute units among compute and communication kernels (\figref{fig:c3_overview}, left), we harness existing direct memory access (DMA) engines on the MI300X GPU and offload communication to them making all compute units available for concurrent GEMM computations (\figref{fig:c3_overview}, right). Given the placement of DMA engines in MI300X GPU (\ssecref{bckg_mi300x}), this has the added advantage of lowering interference in subset of caches as well. To this end, we build concurrent communication collectives (\OPNAMEBOLD) proof-of-concepts (PoCs) for ML collectives such as all-gather and all-to-all\footnote{Note that, as GPU DMA engines do not support arithmetic operations, we do not consider offloading all-reduce kernels to DMAs.}. We demonstrate that our simple POCs are on par with existing communication libraries for bandwidth-bound scenarios while also reducing interference in C3, leading to attainment of on average 72\% of ideal speedup, up to a maximum of 1.67$\times$ speedup. Overall, our work makes a strong case for continued investment and betterment of GPU DMA engines, and we conclude with a discussion of further GPU enhancements to support \PNAME efficiently. 

The key contributions of this work are as follows: 

\begin{itemize}
\item As concurrent computation and communication (\PNAMEBOLD) is an important paradigm in ML, high performance computing (HPC) and other domains, we present a detailed taxonomy and characterization of this paradigm on GPUs. 
\item Our analysis shows that while \PNAME can lead to performance uplifts compared to serial execution, not all of the potential speedup is realized (21\% of ideal speedup is realized). This is not unexpected, as compute and memory interference among concurrent kernels causes slowdowns.
\item To address the above gap, we first observe that \PNAME performance uplifts can be improved (42\% of ideal speedup) by dual strategies of schedule prioritization and careful resource partitioning of compute units on GPUs. 
\item To attain further performance uplifts, unlike current state-of-art communication libraries which offload communication to GPU compute-units, we build concurrent communication collectives (\OPNAMEBOLD)  proof-of-concepts, which offload communication to DMA engines on GPUs. 
\item Using \OPNAME, we demonstrate that \PNAME interference is lowered, considerably closing the gap between realized and ideal speedup (on average 72\% of ideal speedup is realized).
\item Overall, we make a strong case to enhance the capabilities of DMA engines on GPUs, as they can play an important role in efficiently supporting \PNAME in GPUs. 

\end{itemize}

\putsec{bckg}{Background}

\begin{figure}[t]
    \centering
    \includegraphics[scale=0.3,trim=0.1cm 3cm 6cm 0.5cm,clip]{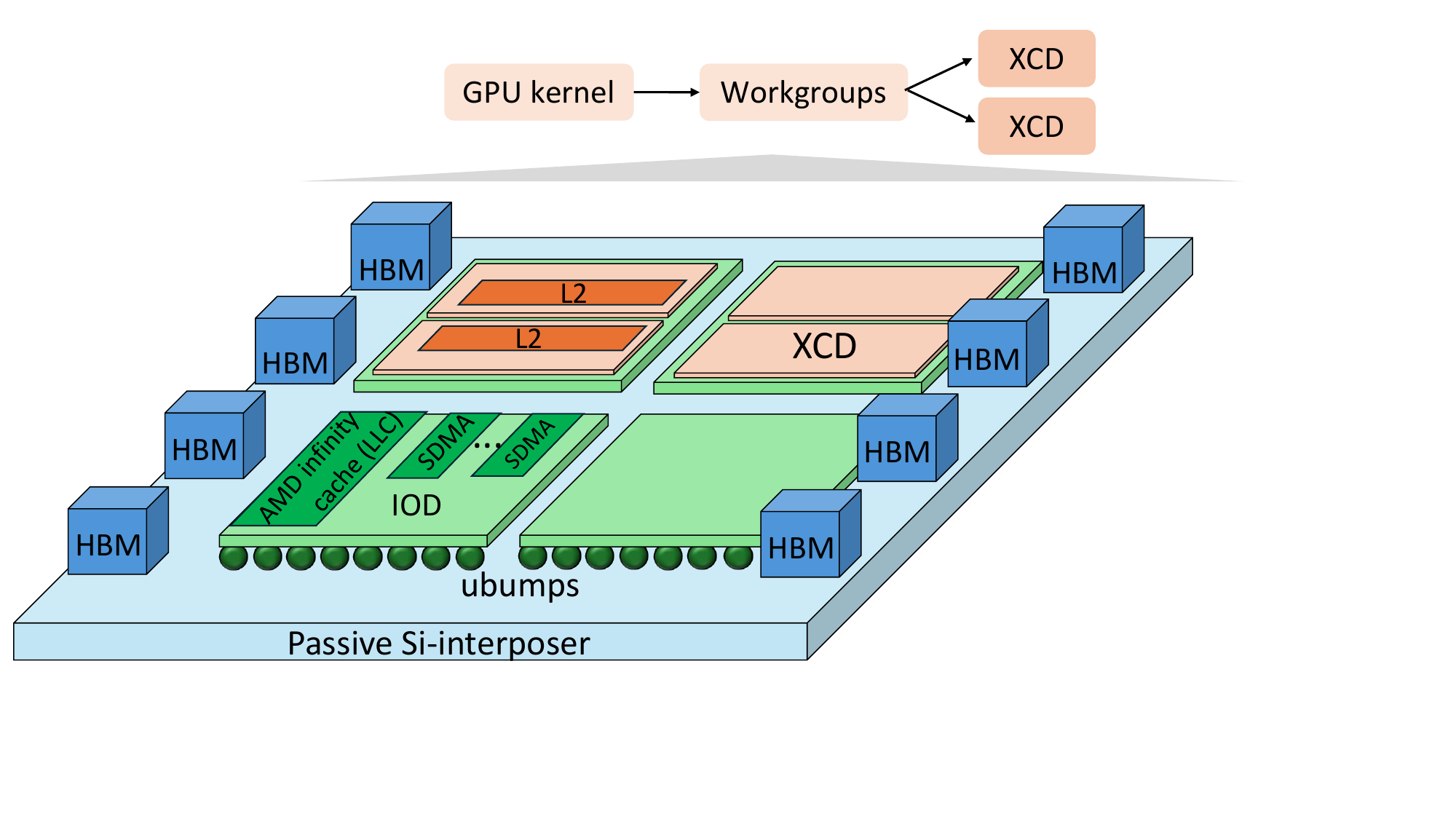}
    \caption{State-of-art AMD Instinct\textsuperscript{\texttrademark} MI300X.}
    \label{fig:bckg_mi300x}
\end{figure}

\putssec{bckg_mi300x}{AMD Instinct MI300X GPU and Compute Orchestration}

In this paper, we study \PNAME using AMD Instinct MI300X GPU depicted in \figref{fig:bckg_mi300x}. A single MI300X GPU employs advanced packaging to integrate heterogeneous chiplets. Specifically, each MI300X is comprised of eight accelerator complex dies (XCD) \cite{mi300x-issca} vertically stacked over four I/O dies (IOD), two XCDs per IOD \cite{mi300x-issca,mi300x-vlsi}. 

The IODs are comprised of AMD Infinity Cache\textsuperscript{\texttrademark}, a shared memory-side last-level cache (LLC) which is 256MB in size. Also, the IODs contain the memory interface to the on-package HBM. Each MI300X has a total of eight HBM stacks, each with 24GB (total of 192GB) for a combined peak memory bandwidth of 5.3TB/s~\cite{mi300x}. Additionally, the IODs also contain 14 DMA copy engines\footnote{AMD HSA runtime API~\cite{hsa} allows querying of available DMA engines on MI300X.}, termed SDMA (or system DMA) which are available for intra-node data transfers between GPUs with necessary address mapping support.

Each XCD is comprised of 38 active compute units (CUs), which are highly threaded and parallel GPU processor cores. Each XCD also has shared L2 cache of 4MB shared across all CUs within the XCD. Overall, in an MI300X GPU, there are 304 CUs. Also depicted in \figref{fig:bckg_mi300x}, is compute orchestration. Computations are offloaded to GPUs as \textit{kernels}, each comprising multiple \textit{workgroups} which are scheduled on available CUs across all XCDs. 

Large-scale ML (the focus of this work) often employs multiple GPUs in tandem. In this work, we focus on the AMD MI300X Infinity Platform comprised of a 8x MI300X node with a fully-connected topology. Each MI300X connects to seven other MI300X GPUs using AMD Infinity Fabric\textsuperscript{\texttrademark}~\cite{mi300x} bi-directional links (each with uni-directional bandwidth of 64GB/s/link). 

\putssec{bckg_dma}{Direct Memory Access (DMA) Engines in GPUs}

\begin{figure}[t]
    \centering
    \includegraphics[scale=0.4,trim=3cm 7cm 5cm 3cm,clip]
    {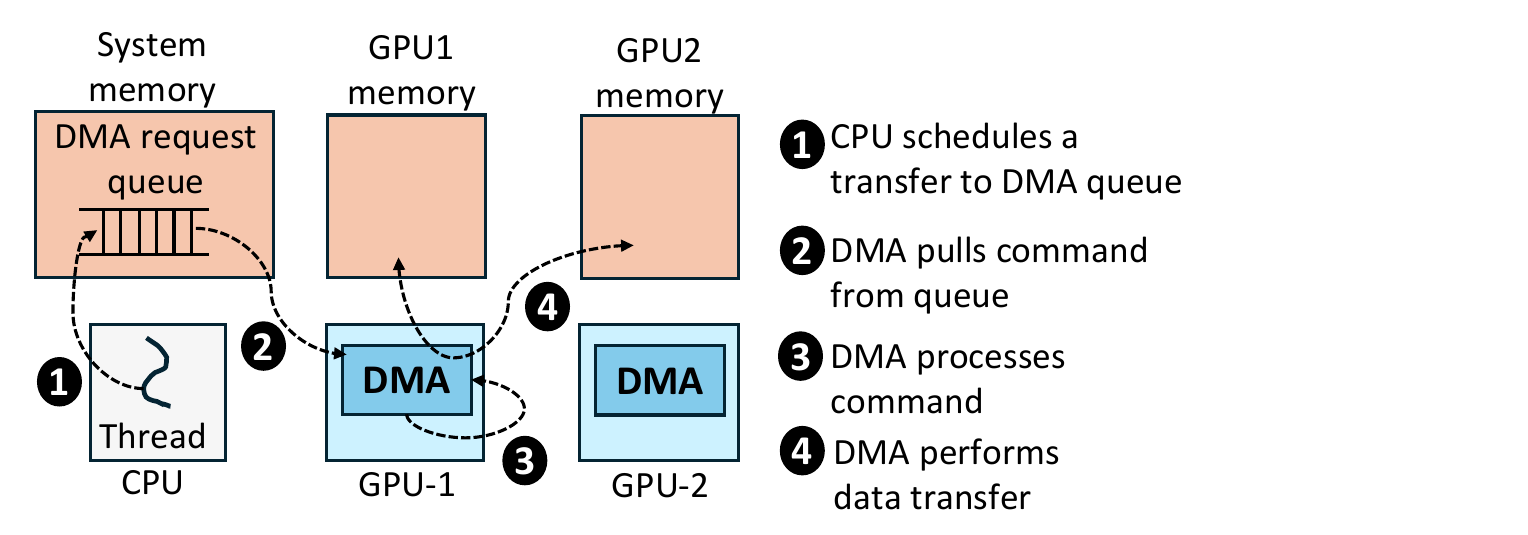}
    \caption{Offloading a data-transfer to DMA in MI300X.}
    \label{fig:c3_dma_bckg}
\end{figure}

As discussed above, each MI300X is comprised of 14 SDMA copy engines available for transfers between address-space shared GPUs. We depict the steps involved for users to use these DMA engines in \figref{fig:c3_dma_bckg}. Using either heterogeneous-computing interface for portability (HIP)~\cite{hip} or heterogeneous system architecture (HSA)~\cite{hsa} runtime API calls, at user-level, a programmer requests a single data transfer to be done using SDMA engines. Under the hood, this causes the runtime on CPU to place a command packet (\circled{1}) in the DMA queue  placed in system memory for a specific GPU (either source or destination GPU for the transfer). 
The GPU DMA engine gets notified and fetches the command from the queue (\circled{2}) and processes it (\circled{3}). Once decoded, the DMA engine issues necessary reads/writes from/to HBM memory of source/destination GPU respectively to complete the transfer (\circled{4}). 

\putssec{bckg_c3}{ML Operators and \PNAME in ML}
As large-scale ML training and inference increasingly rely on GPU clusters, GPUs need to support both efficient computation and efficient communication. Furthermore, there are several algorithmic choices in ML that lead to \textit{concurrent} computation and communication. \PNAME entails a pair of computation kernel and communication kernel that have no data dependencies and as such as be scheduled concurrently on a given GPU. 

Examples of such algorithmic choices include data-parallelism~\cite{dataParallel90}, FSDP~\cite{zhao2023fsdp}, nano-batching~\cite{zhu2024nanoflow} and more. In data parallelism, the backpropagation phase of ML training comprises matrix-matrix multiplication gradient kernels (both input and weight gradients), which can be scheduled concurrently with reduction (communication) of weight gradients across GPUs for intra-layer concurrency (input gradient) or inter-layer concurrency (weight gradient of previous layer). Similarly, techniques like FSDP, gather model weights for a given layer on a GPU (communication), while performing computations of previous layers. Finally, optimizations like nano-batching~\cite{zhu2024nanoflow} break down a single batch of inputs into nano-batches opening up opportunity for compute and communication kernels from different nano-batches to be co-scheduled on a GPU. Overall, given the ample prevalence of \PNAME in ML, characterization and optimization of \PNAME on GPUs is important and hence the focus of this work. 

\putsec{taxonomy}{\PNAME Taxonomy}

We begin with a taxonomy for \PNAME and subsequently anchor our characterizations and optimizations on this taxonomy. 

\begin{figure}[!t]
    \centering
    \includegraphics[scale=0.42]{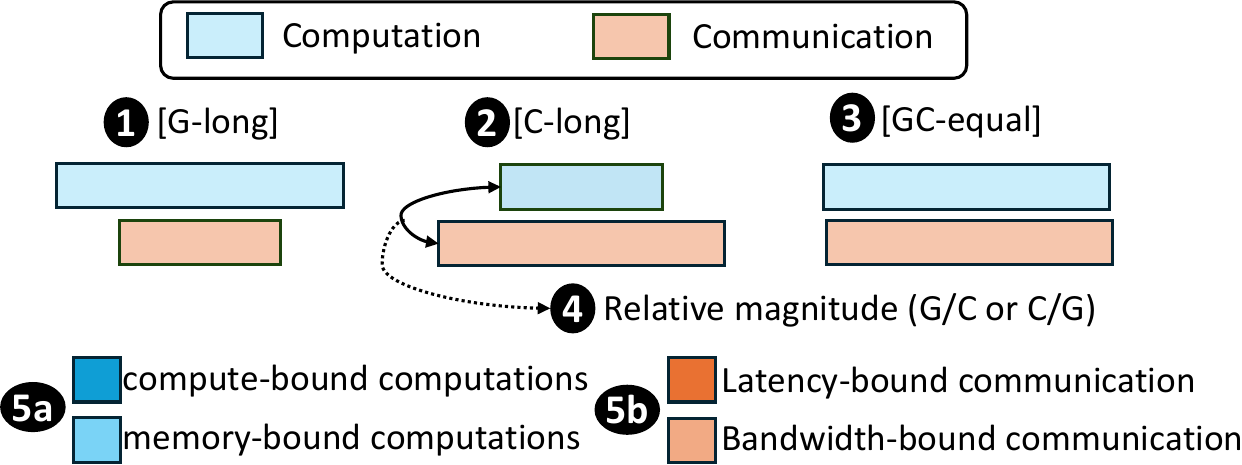}
    \caption{\PNAME taxonomy.}
    \label{fig:c3_taxonomy}
\end{figure}

As discussed in \ssecref{bckg_c3}, the focus of our work is on the manifestations of \PNAME in ML. More specifically, while there exist a variety of operators/computations in ML, in this work we focus on two primary operators, namely general matrix-matrix multiplications (or GEMM) kernels and communication kernels, as they contribute to the majority of the ML execution time in both training and inference scenarios across a spectrum of models~\cite{patiToTC23}. Based on this, we depict our proposed \PNAME taxonomy in \figref{fig:c3_taxonomy}.
Specifically, we consider three key types of \PNAME: \circled{1} \textbf{G-long}, \circled{2} \textbf{C-long} and \circled{3} \textbf{GC-equal}.

We use execution times in isolation for our taxonomy. Thus, G-long is a \PNAME manifestation where GEMM time in isolation is \textgreater 115\% of communication time. Similarly, C-long implies communication time in isolation is \textgreater 115\% of GEMM time. Finally, GC-equal is a \PNAME manifestation where both kernels are comparable (within 15\% of each other).

Even within this framework, we further observe that the relative magnitude (\circled{4} in Figure~\ref{fig:c3_taxonomy}) of computation and communication kernels with respect to each other has an effect on expected interference incurred and as such performance attained. That is, when two kernels run concurrently on a GPU, they interfere with each other in both compute (splitting of compute units) and memory subsystem (caches, HBM). As such, we consider this part of our taxonomy and vary the relative magnitudes of the \PNAME scenarios under study in our work. 

Finally, not all GEMM or communication kernels are the same. For GEMMs, we consider two broad categories of \textit{compute-bound} and \textit{memory-bound} (\circled{5a} in Figure~\ref{fig:c3_taxonomy}) kernels. We define a kernel to be compute-bound if its measured op-to-byte ratio is larger than machine op-to-byte as calculated from peak compute and memory throughput of underlying processor (kernel is memory-bound otherwise). For communication, we consider two types of commonly used multi-GPU ML collective kernels - \textit{all-gather} and \textit{all-to-all}. Similar to GEMM kernels, such ML collective kernels, depending on communication size, can be \textit{latency-bound} or \textit{bandwidth-bound} (\circled{5b}). We classify a communication kernel with its associated size as latency-bound if the kernel latency at/before this size does not increase commensurate to size. We further discuss the specific sizes and \PNAME manifestations that we consider to cover this taxonomy, the source (ML model) for these manifestations and rationale in \ssecref{methodology}.

\putsec{c3_baseline}{Baseline \PNAME Characterization}

\putssec{methodology}{Methodology}

\putsssec{meth_system}{System Setup}
As discussed in \ssecref{bckg_mi300x}, we study \PNAME in ML using the AMD MI300X Infinity Platform comprised of 8x MI300X node with a fully connected topology.

Further, recall that we focus on two primary ML operators, namely, GEMM kernels and for communication, ML collectives kernels such as \textit{all-gather} and \textit{all-to-all}. For former, we employ AMD ROCm\textsuperscript{\texttrademark}~\cite{rocm} rocBLAS library~\cite{rocblas} comprised of high-performance GEMM kernels. For ML collectives, we employ AMD ROCm Communication Collectives Library  or RCCL~\cite{rccl}, a library of standard collective communication routines for GPUs. 

We use multiple GPU streams~\cite{hipstream} or independent set of GPU kernels that can be co-scheduled on the GPU to concurrently launch GEMM (computation) and ML collective (communication) kernels by scheduling each type of kernel in its independent stream. Additionally, we leverage the feature available on MI300X GPU to reserve CUs for specific stream to study compute needs of kernels. We use rocprof~\cite{rocprof} GPU kernel performance profiling tool to assess kernel execution times for this study. We run total of 15 executions, first 6 are warm ups and then 9 are actual measured.

\putsssec{meth_gemm_comm_sizes}{\PNAME Manifestations Under Study}

\begin{table}[t]
\centering
\tiny
\caption{Computations (GEMMs) studied, tags and source.}
\label{tab:c3_gemms}
\resizebox{\columnwidth}{!}{%
\begin{tabular}{|l|l|l|l|l|l|l|}
\hline
\textbf{gemm-tag} & \textbf{gemm-size} & \textbf{source} \\ \hline
cb1 & 8192x8192x8192 & LLaMA-70B \\ \hline
cb2 & 16384x8192x16384 & LLaMA-405B \\ \hline
cb3 & 16384x16384x8192 & LLaMA-405B \\ \hline
cb4 & 18432x8192x16384 & LLaMA-405B \\ \hline
cb5 & 106496x8192x16384 & LLaMA-405B \\ \hline
mb1 & 8192x57344x8192 & LLaMA-70B \\ \hline
mb2 & 16384x106496x8192 & LLaMA-405B \\ \hline
\end{tabular}%
}
\end{table}

We discuss the GEMM kernels we study in this work, their sizes and their sources in \tabref{tab:c3_gemms}. As listed, we source our GEMM sizes from training of LLaMA-70B and LLaMA-405B~\cite{llama3-24} models processing 8192 tokens (i.e., product of input length and batch-size) in a given iteration. Further, for ease of reading, we tag these GEMM sizes/kernels as compute-bound (\textbf{cb}) or memory-bound (\textbf{mb}). Recall that, we define a kernel to be compute-bound if its measured op-to-byte ratio is larger than machine op-to-byte as calculated from peak compute and memory throughput of underlying processor (kernel is memory-bound otherwise).

We list the \PNAME manifestations we study in this work in \tabref{tab:c3_roofline}. We first provide a tag for \PNAME being studied using GEMM type followed by the array/data size of the parallel collective operation. This is listed in column \textbf{C3} in the table. Further, in all our analysis, we separately present the results for different collective types (all-gather, all-to-all). We also list in the table the \textbf{source} for each \PNAME manifestation. As shown, of 15 unique \PNAME combinations we have, seven are manifested in training of LLaMA-70B and LLaMA-405B~\cite{llama3-24} models (we assume 8-way sharding and FSDP~\cite{zhao2023fsdp}) using all-gather collective. To provide good coverage for our proposed taxonomy in \secref{taxonomy}, we add additional synthetic \PNAME manifestations, wherein, we keep the GEMM kernel size as observed in LLaMA models (\tabref{tab:c3_gemms}) but add more communication sizes and repeat all \PNAME scenarios for all-to-all collective as well. 

\begin{table}[t]
\centering
\caption{C3 combinations considered and taxonomy.}
\label{tab:c3_roofline}
\resizebox{\columnwidth}{!}{%
\begin{tabular}{|l|l|l|l|}
%\hline
\multicolumn{4}{l}{\textbf{C3-type: G-long}} \\ \hline
\textbf{C3} & \textbf{source} & \textbf{C3} & \textbf{source}\\ \hline
mb1\_896M & LLaMA-70B & mb2\_3.25G & LLaMA-405B \\ \hline
mb1\_4G & synthetic & mb1\_6G & synthetic \\ \hline
cb3\_512M & LLaMA-405B & cb4\_512M & LLaMA-405B \\ \hline
cb5\_1.63G & LLaMA-405B &  cb4\_1G & synthetic \\ \hline
\multicolumn{4}{l}{\textbf{C3-type: C-long}}\\ \hline
mb1\_13G & synthetic & cb2\_3.25G & LLaMA-405B \\ \hline
cb4\_2.5G & synthetic & cb1\_896M & LLaMA-70B \\ \hline
cb5\_20G & synthetic  \\ \hline
\multicolumn{4}{l}{\textbf{C3-type: GC-equal}}\\ \hline
mb2\_26.5G & synthetic & cb5\_13G & synthetic \\ \hline
\end{tabular}%
}
\end{table}

We also list the taxonomy for each \PNAME in \tabref{tab:c3_roofline}. Notice that we have more G-long scenarios than C-long scenarios than GC-equal scenarios. This is so, as shown in \tabref{tab:c3_roofline}, majority of \PNAME manifestations in LLaMA models today are of G-long type and we consciously wanted to limit the synthetic scenarios we added. However, we provide all our results grouped by \PNAME type to understand the performance for each of these different types. 

\putssec{isolated}{Isolated Execution Characterization}
\begin{figure*}[!th]
    \centering
    \includegraphics[width=\linewidth]{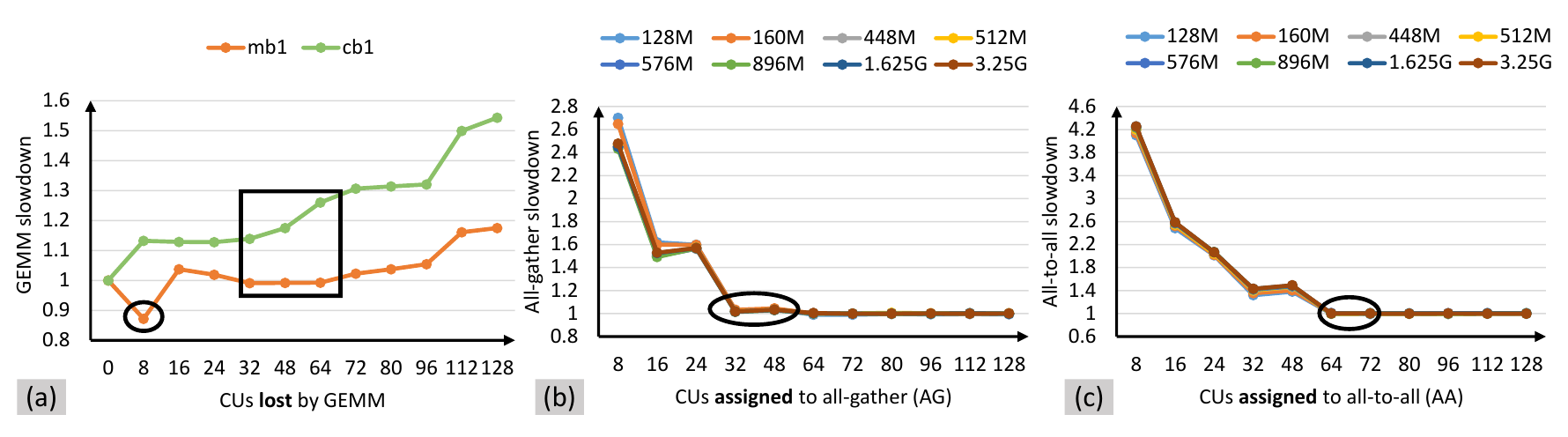}
    \caption{(a) GEMM kernel slowdown due to loss of compute units (CUs) in GPU. (b) All-gather, (c) All-to-all kernel slowdown with specific \# CUs assigned vs. default CUs (All-gather default \#CUs=64, All-to-all default \#CUs=56). For single partition MI300X with eight XCDs, eight is the minimum number of CUs that can be assigned to a kernel.}
    \label{fig:isolated_cu_slowdown}
\end{figure*}

We first profile the computation and communication kernels in isolation to ascertain their compute and memory needs and assess potential for performance when they are executed concurrently. 

\putsssec{isolated_compute}{Compute Needs} Recall that, to execute computation and communication concurrently on GPUs, we launch two independent kernels and as such available compute cores or compute units (CUs) for MI300X (\ssecref{bckg_mi300x}) are split between these two kernels. This compute interference can slowdown each of these kernels. We study this in \figref{fig:isolated_cu_slowdown} (a) for GEMM kernels and (b)/(c) for subset of all-gather and all-to-all sizes (rest of the sizes show similar behavior). 

For GEMM kernel slowdown, we depict two of the available seven GEMMs under study (\tabref{tab:c3_gemms}) in Figure~\ref{fig:isolated_cu_slowdown}(a), as they represent the extremes in terms of slowdown. To plot slowdown, we compare a GEMM execution when the specified number of CUs are taken away from the kernel to an execution where all CUs are available to the kernel (0 on x-axis is when the kernel has all 304 CUs available to it). As depicted in the figure, memory-bound GEMMs are resilient to CU loss, even attaining speedups\footnote{We observe better cache behavior for this GEMM due to less number of concurrent threads.} (highlighted with a circle), while compute-bound GEMMs can incur increasing slowdowns as more CUs are taken away from the GEMMs. As for communication, we do a similar slowdown analysis in Figure~\ref{fig:isolated_cu_slowdown}(b)/(c) and observe that unlike GEMMs, all-gather kernels need 32 CUs, while all-to-all kernels need 64 CUs beyond which there is no benefit to allocating more CUs to the kernel.

Overall, considering the compute needs of these kernels in tandem, it can be expected that memory-bound GEMMs can co-exist with communication kernels, because memory-bound GEMMs are resilient to loss of about 32-64 CUs needed by communication kernels. On the other hand, for the same CU loss, compute-bound kernels suffer up to 17-27\% slowdowns.

\putsssec{isolated_memory}{Memory Needs}
\begin{figure}
     \centering
    \includegraphics[scale=0.65]{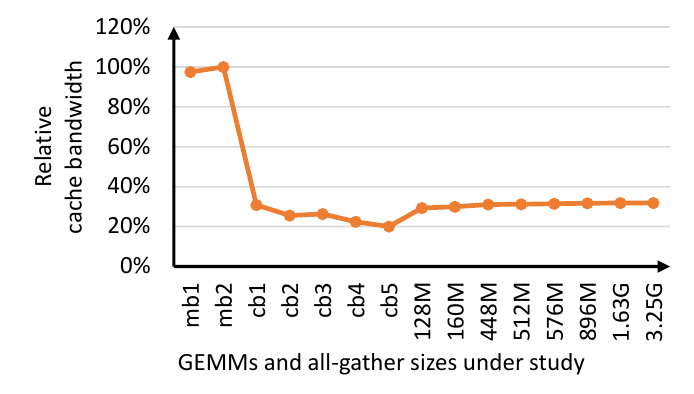}
     \caption{Relative AMD Infinity Cache\textsuperscript{\texttrademark} bandwidth utilization.}
     \label{fig:c3_mall_bw}
\end{figure}

Similar to compute needs analysis above, we next consider memory needs of these kernels. As discussed in \ssecref{bckg_mi300x}, MI300X GPU employs a shared last-level AMD Infinity Cache which, being memory-side, subsumes all of HBM traffic. As such we study the memory bandwidth needs of kernels under study by providing their relative AMD Infinity Cache bandwidth utilization in \figref{fig:c3_mall_bw}. Note that, we only show all-to-all kernels and skip all-gather kernels as the latter have about 14\% lower bandwidth vs. all-to-all kernels.

As depicted in \figref{fig:c3_mall_bw}, the bandwidth needs of memory-bound GEMM kernels dwarf that of all other kernels under study. That said, the compute-bound GEMM and communication kernels are in similar ballpark of bandwidth needs. Further, given that the combined bandwidth needs of compute-bound GEMM kernels and communication kernels still leaves headroom in peak bandwidth available, these kernels can share available bandwidth in concurrence. 

\begin{figure}
     \centering
    \includegraphics[scale=0.5]{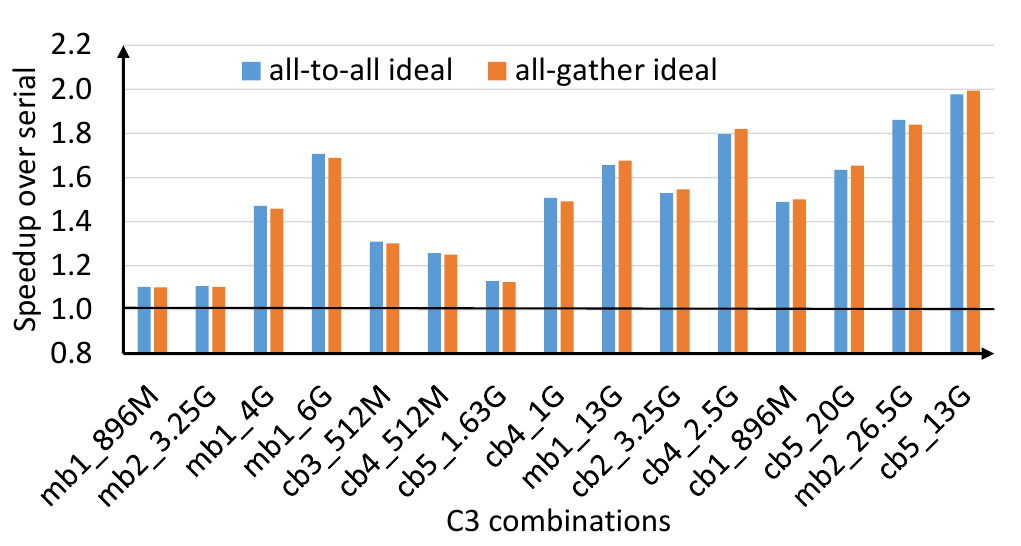}
     \caption{Ideal speedup possible for \PNAME scenarios under study.}
     \label{fig:c3_roofline}
\end{figure}

\putsssec{isolated_roofline}{Ideal speedup projection} Based on above isolated execution analysis of compute and memory needs, we project in \figref{fig:c3_roofline} the ideal speedup possible for \PNAME scenarios under study. For this analysis, we assume that the best speedup possible is when the smaller of the two kernels (computation or communication) happens completely in the shadow of the other. As depicted, wide variety of speedups are possible (1.1$\times$ to close to 2$\times$) and this is largely dictated by relative magnitudes of these kernels (\secref{taxonomy}). However, based on compute/memory needs analysis, due to compute/memory interference not all of this ideal performance can be attained. 

Further, note that, based on runtime decisions or GPU-GPU execution variation or kernel-size/type, different degrees of overlap can manifest, resulting in different ideal speedups. To precisely tackle this we design \PNAME taxonomy so that we can consider such varied manifestations and analyze our proposals across these different \PNAME manifestations.

\putssec{c3_base}{\PNAME Characterization}

\begin{figure}[t]
    %\centering
    \includegraphics[scale=0.4]{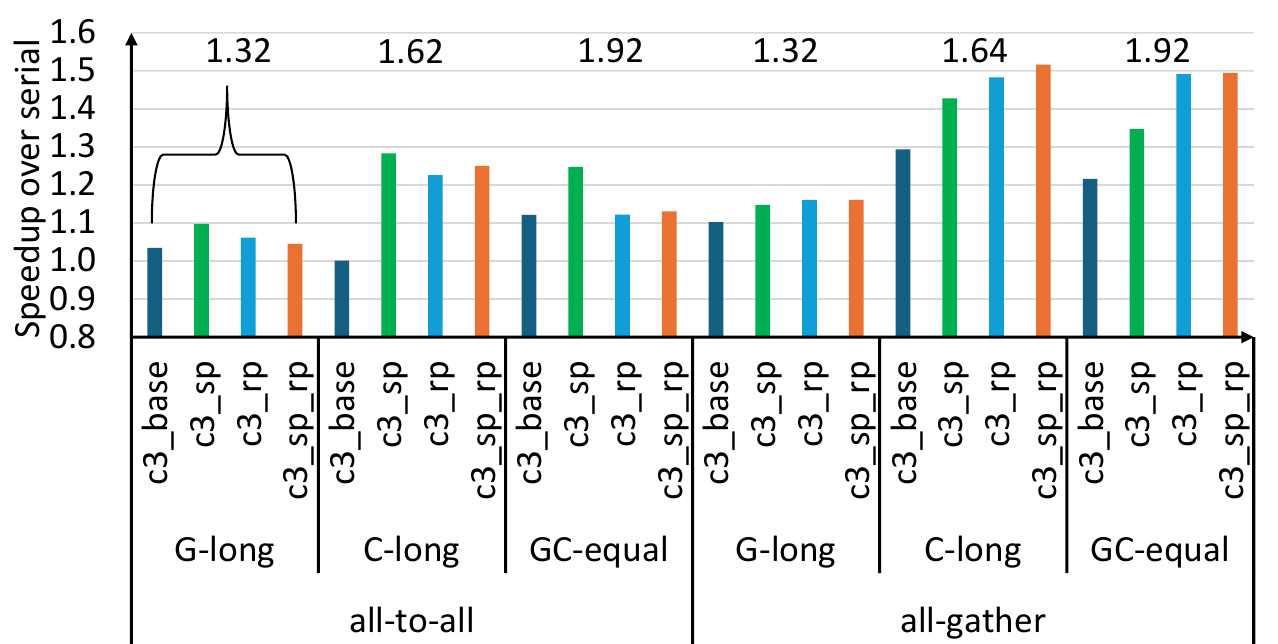}
    \caption{Speedups for \PNAME scenarios under study with and without schedule prioritization and resource partitioning.}
    \label{fig:c3_base}
\end{figure}

We present characterization of \PNAME for different scenarios under study in \figref{fig:c3_base}. In this figure we refer to baseline \PNAME performance as \textbf{c3\_base}. Note that, with concurrent scheduling, we schedule GEMM kernel first (as scheduling both the kernels at the exact same time is challenging, we minimized the scheduling delay between the two through code optimizations) in \textbf{c3\_base} executions. 

In \figref{fig:c3_base}, we present average speedups for different groupings of \PNAME scenarios (collectives, and taxonomy). We also show in the graph the ideal speedups possible (marked at the top of the graph). The figure shows that \textbf{c3\_base} speedups range from no speedups to up to 1.3$\times$ speedups. On average, \textbf{c3\_base} attains  1.13$\times$ speedup which is about 21\% of ideal speedup. Note that, this is not unexpected. In fact, prior work~\cite{hwang2023ark} observed slowdowns with \PNAME due to mutual interference of concurrent kernels.

Further, all-to-all attains about 0-13\% of ideal speedups while all-gather attains about 24-46\%. The slight edge for all-gather is due to the fact the overall memory traffic (hence memory interference) and compute needs are lower for all-gather vs. all-to-all (\sssecref{isolated_compute}). To understand lower memory traffic of all-gather, in simplistic terms, in all-gather, each GPU begins with a single data buffer and the end state is that every GPU holds the complete aggregated data from all other GPUs. In contrast, with all-to-all, each GPU starts with distinct data buffer for every other GPU (lot more data) and concludes with each GPU receiving unique data from all other GPUs (effectively a transpose of data buffers).

\putsec{c3_base_opt}{Optimizing \PNAME Performance}
Next, to improve \PNAME performance reported above, we evaluate dual strategies of schedule prioritization and careful resource partitioning of compute units on GPUs.

\putssec{schedule_prio}{Schedule Prioritization}
Note that, thus far, for concurrent scheduling, we chose to schedule GEMM kernel first. But based off our compute needs analysis, we observe that communication kernels have far lower CU needs as compared to GEMM kernels (\sssecref{isolated_compute}). As such, if GEMM kernels are scheduled first, the internal GPU scheduler can in some cases allocate majority of CUs to GEMM kernel leading to potential starvation. In contrast, if we employ, \textit{schedule prioritization}, that is, schedule communication kernel first, later scheduled GEMM kernel will definitely have compute resources to be allocated. Overall, the key insight of schedule prioritization is that prioritizing or providing quality of service for kernel with smaller and complimentary resource requirement helps concurrent performance.

To realize schedule prioritization, from CPU-side, we first schedule communication kernel in a stream and then immediately after, schedule GEMM kernel in concurrent stream. We present the results of schedule prioritization, referred in \figref{fig:c3_base} as \textbf{c3\_sp} and show the speedups attained. Across all \PNAME types and collectives, we observe that schedule prioritization improves the speedups attained for \PNAME. Overall, instead of 0-13\%  of ideal speedup, for all-to-all, \textbf{c3\_sp} attains 27-46\% instead. Similarly, instead of 24-46\% of ideal speedup, for all-gather, \textbf{c3\_sp} attains 38-67\% of ideal speedup. Overall, on average, \textbf{c3\_sp} helps attain 42\% 
of ideal speedup (up from 21\% of \textbf{c3\_base}).

\putssec{resource_part}{Resource Partitioning}
Note that schedule prioritization is not the only way to prevent starvation of kernel with lower resource needs. We also observe that we can leverage the feature available on MI300X GPU to reserve compute units (CUs) for specific stream to more deterministically allocate CUs to communication kernels and attain similar benefits. 

We show this in \figref{fig:c3_base} as \textbf{c3\_rp} which adds resource partitioning (rp) to \textbf{c3\_base}. We sweep all possible powers-of-two CU allocations for communication kernels (and consequently take CUs away from GEMM kernel) and plot the best performing one as \textbf{c3\_rp}. As depicted, \textbf{c3\_rp} delivers performance improvements over \textbf{c3\_base} which is almost the performance delivered by \textbf{c3\_sp} (41\% of ideal speedup). Note, that, adding resource partitioning to \textbf{c3\_sp}, depicted as \textbf{c3\_sp\_rp}, did not improve performance any further. 

\putssec{runtime_hints}{Runtime Heuristics for Improved \PNAME Performance}
We believe that either schedule prioritization or resource partitioning can help improve \PNAME performance, while former is indeed simpler to implement. To this end, we provide some heuristics that can guide a runtime while employing them. 

\textbf{Heuristic for Schedule Prioritization:} As runtimes launch GPU kernels, they can use the information about number of workgroups per kernel (\ssecref{bckg_mi300x}) as a proxy for CU requirement of a kernel. Using this information, excepting any other prioritization that supersedes, the runtime can employ scheduling order in the order of resource requirements  (number of workgroups), low to high.

\textbf{Heuristic for Resource Partitioning:} For resource partitioning across concurrent kernels, we describe a runtime heuristic to allocate CUs across concurrent kernels. To do so, we picked one memory-bound GEMM kernel, one compute-bound GEMM kernel, and two all-gather and all-to-all collective sizes (latency-bound, bandwidth-bound) and used the compute need analysis done in \sssecref{isolated_compute} to build a lookup table of potential slowdowns when the kernel loses a given number of CUs. Note that, for a given GPU this is to be done once. 

Next, for any \PNAME scenario, we scale roofline GEMM and communication times by these slowdowns for different number of CUs to identify the CU allocation that leads to best performance, that is max(GEMM, communication) time is lowest. For roofline times, we simply focus on peak compute, memory and network throughputs and assume 70\% efficiency (taking the average of our/observed~\cite{mi300x-perf} peak compute, memory and network efficiencies). We observe that this simple heuristic, predicts CU allocation necessary for 24 of 30 \PNAME scenarios. For the rest, in comparison to sweeping all possible CU allocations, our heuristic, at best loses 1.5\%.
\putsec{c3_conccl}{Optimizing \PNAME with \OPNAME}

\begin{figure}[t]
\centering
 \includegraphics[scale=0.5]{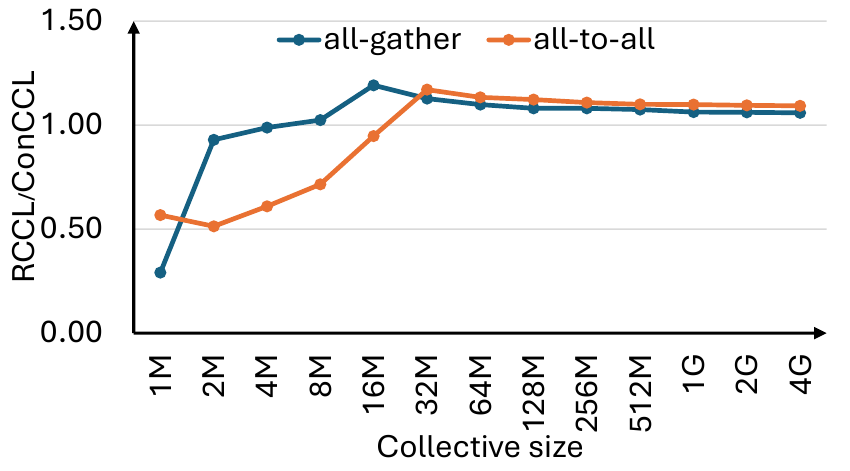}
 \caption{\OPNAME speedup over CU-based collective (RCCL).}
 \label{fig:conccl_isolation}
\end{figure}

\putssec{conccl}{Motivation for DMA Offloads of Communication}
For co-scheduled computation and communication, the performance limiters are the compute and memory interference incurred by sharing of compute and memory resources between concurrent kernels. That is, a GEMM kernel in isolation could have been allocated all of 304 compute units (CUs) available on MI300X. In \PNAME , however, number of CUs allocated will be lower causing what we term as compute interference and leading to lower speedups. Similar interference also occurs in the memory sub-system (e.g., caches, HBM). 

One way to tackle compute interference completely and memory interference partially, is to offload communication to DMA engines on MI300X. By doing so, we can free up all available compute units for concurrent computation kernel. Further, as DMA engines are placed in IOD beyond L2 caches (\ssecref{bckg_dma}), this also eliminates any L1/L2 portion of memory interference. These dual benefits motivate us to offload communication to DMA engines on MI300X.

\putssec{conccl_dma}{\OPNAME: DMA-collectives PoCs}
With the above strong motivation, we build \OPNAMEBOLD proof-of-concepts (PoCs), which are DMA-based collectives wherein we offload the collectives to DMA engines on MI300X. As MI300X DMA engines do not expose any computational functionality, we only build these PoCs for all-gather and all-to-all (and not all-reduce collective which involves reduction). 

We harness the fully-connected topology of MI300X to keep the design of collectives simple. That is, we break down the collective operation into a series of individual transfers (going to different GPUs) and we schedule each such transfer on a specific available DMA engine on MI300X (recall that 14 engines are available). We leverage AMD HSA API call \texttt{hsa\_amd\_memory\_async\_copy\_on\_engine}~\cite{rocr-runtime} to schedule a single transfer at a given engine. Also, we implement a simple direct algorithm for our collectives. As an example, for all-gather this means that our collective implementation comprises a single step: every GPU reads the data buffer it owns and writes it to every other GPU. Additionally, unlike RCCL communication kernels which are orchestrated by launching GPU kernels, as discussed in \ssecref{bckg_dma}, DMA engine transfers are orchestrated by CPU. 

\begin{figure}[t]
\centering
 \includegraphics[width=\columnwidth]{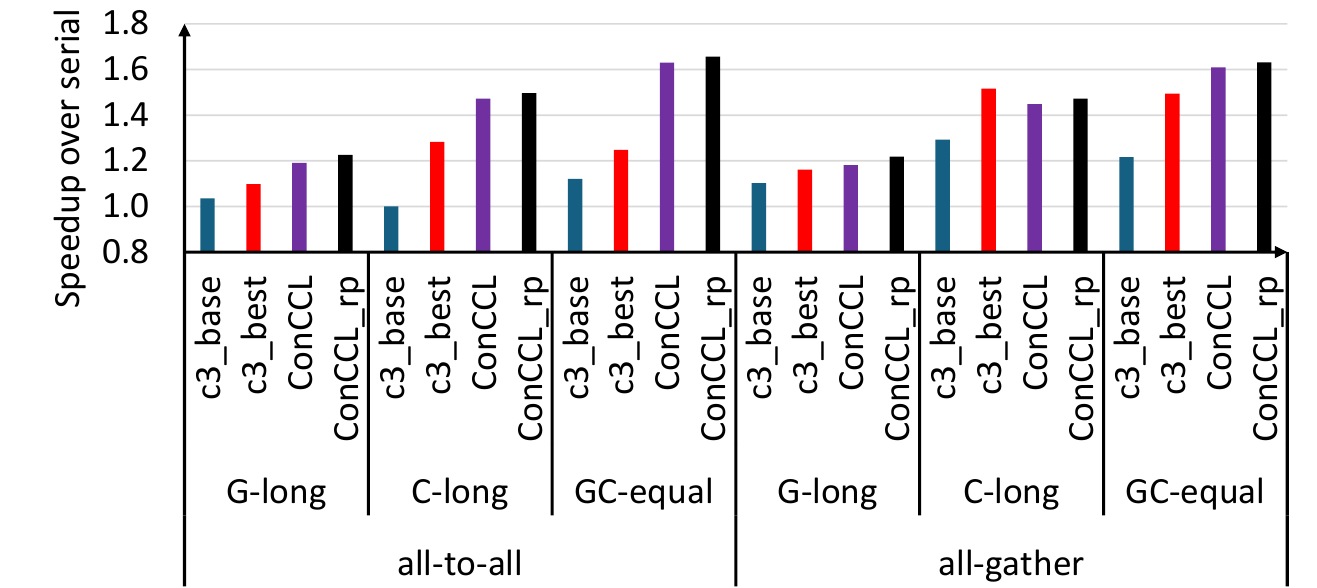}
 \caption{\PNAME speedup with \OPNAME.}
 \label{fig:conccl_speedup}
\end{figure}

Finally, we term \OPNAME as PoCs for our goal in this work is not to build a high-performant communication collectives library but to use these PoCs to analyze if offloading concurrent communication collectives to DMA engines helps improve speedups possible with \PNAME. We leave building a high-performance DMA-based collectives library to future work. 

\putssec{isolated_conccl}{Isolated \OPNAME Characterization}
We first provide a comparison of our PoCs to state-of-art RCCL communication library on MI300x. We show this comparison in \figref{fig:conccl_isolation}. We follow the same setup details as outlined in \sssecref{meth_system} and model multiple warm-up executions before actual measured executions.

As shown in \figref{fig:conccl_isolation}, our simple PoCs are slower than RCCL library for \textless32MB collective size by as much as 4$\times$. This is so, as it can be costly to both launch transfers from CPU to DMA engines on GPU and synchronize with CPU once the transfers are done. This launch/sync cost is not amortized for smaller sizes. That said, for larger sizes, our proposed all-gather PoCs is at par with RCCL library. Finally, recall that, as the smallest communication size we consider in our \PNAME scenarios is 128MB (\tabref{tab:c3_roofline}), evaluating \PNAME with RCCL or \OPNAME is a fair evaluation as in this region the performance of both is at-par. 

\putssec{c3_conccl}{\PNAME with \OPNAME Characterization}
Next, we evaluate the speedups of \PNAME with \OPNAME over sequential execution. We depict this in \figref{fig:conccl_speedup}. 

We first compare \textbf{c3\_base} configuration, which is \PNAME without schedule prioritization or resource partitioning, to \OPNAME without such optimizations. As shown, across the board, \OPNAME delivers higher speedups in comparison to \textbf{c3\_base}. This is attributed to lowering the compute interference incurred by GEMM kernels in \textbf{c3\_base}. Additionally, GEMM kernels also do not incur any memory interference in any higher level of caches such as L1/L2 with DMA offloads.

Overall, while \textbf{c3\_base} attains 21\% of available ideal speedup, with DMA offloads, \OPNAME attains 66\% of ideal speedup. Further, \OPNAME benefits are even more pronounced for all-to-all (\textbf{c3\_base}: 1.05$\times$,  \OPNAME: 1.43$\times$) which has higher compute needs (64 CUs) and as such incurs more compute interference and also has higher memory traffic/interference in higher level caches. 

\putssec{conccl_sp}{Schedule prioritization with \OPNAME}
Unlike \textbf{c3\_base}, where both computation and communication are scheduled on GPU CUs making schedule prioritization important, with \OPNAME, as communication is scheduled to DMA engines and computation to CUs (two separate entities), schedule prioritization is not necessary for \OPNAME. 

\putssec{conccl_rp}{Resource partitioning with \OPNAME}
Finally, we consider resource partitioning for \OPNAME. Since \OPNAME does not use any compute units for communication, the allocation decision here is different from \textbf{c3\_base}. Here we observe that, \textit{only} for memory-bound GEMM kernels, as depicted in \figref{fig:isolated_cu_slowdown}, taking CUs away from GEMM kernel can lead to performance improvement due to improved cache behavior. Such improvement can also aid in \PNAME runs where communication is offloaded to DMA engines. 

To study this, only for memory-bound GEMM kernels, we create a variant of \OPNAME, \textbf{ConCCL\_rp} and depict its performance in \figref{fig:conccl_speedup}. This variant performs slightly better than \OPNAME. Point to note that, both \OPNAME and \textbf{ConCCL\_rp} performs significantly better than \textbf{c3\_best} variant (best of all \textbf{c3} variants in \secref{c3_baseline}). We see here that on average, while \textbf{c3\_best} attains 48\% of ideal speedup, \OPNAME attains 66\% and \textbf{ConCCL\_rp} attains 72\% of ideal speedup, considerably closing the gap between attained and ideal speedup.

\putssec{conccl_rp_runtimes}{Runtime Heuristic for Resource partitioning with \OPNAME}
For resource partitioning with \OPNAME, we recommend a much smaller subset of steps as proposed in \ssecref{runtime_hints} for baseline \PNAME. Specifically, only the CU-loss slowdown table for a memory-bound GEMM kernel ought to be created. This table helps the runtime identify the necessary number of CUs to remove from concurrent memory-bound GEMM kernel to maximize speedup. In our analysis, for MI300X, taking away eight CUs lead to speedups for memory-bound GEMM kernels.
\putsec{discussion}{Discussion}

\putssec{future}{System Evolution for Efficient \PNAME}
While \OPNAME leads to more efficient \PNAME on GPUs, we believe additional system evolution can further help in this regard. We highlight some potential techniques below. 

\putsssec{mem_interference}{Addressing Memory Interference in \PNAME}
Our approach mitigates compute interference by offloading communication to the GPU's DMA engines. On the MI300X GPU, DMA transfers inherently do not interact with the L1 and L2 caches, effectively eliminating cache interference at these levels. That said, contention for HBM bandwidth remains, impacting performance. We leave to future work exploration of techniques such as memory-bandwidth partitioning via memory space (channel) partitioning amongst kernels, memory-aware scheduling~\cite{pati2024t3} to explicitly tackle memory interference. Specifically for latter, \PNAME taxonomy hints to hardware can aid in memory traffic prioritization. 

\putsssec{conccl_ar}{Accelerating \PNAME with All-Reduce Collective}
Our \OPNAME PoCs focus on offloading all-gather and all-to-all collectives to DMA engines. However, all-reduce operations involve both communication and computation (e.g., summing values across GPUs) and since DMA engines do not currently support arithmetic operations we could not offload all-reduce collectives to DMA engines. That said, all-reduce is also used in ML and offloading this to DMA engines can be useful. Investigating addition of arithmetic units to DMA engines is a possibility although the area/power costs of doing so ought to be balanced with possible returns. Alternately, a hybrid approach can be followed. That is, as all-reduce is comprised of a reduce-scatter and all-gather operation, only for latter DMA engines can be harnessed.

Further, we believe that our proposed techniques of schedule prioritization and resource partitioning are also applicable to all-reduce. This is so, as the key insight of the techniques is that prioritizing or providing QoS for kernel with smaller/complimentary resource requirement helps concurrent performance. As all-reduce has this behavior just like all-gather/all-to-all vs. GEMMs (low CU needs, etc.), these techniques can be applicable to all-reduce as well.

\putssec{future_other}{Other considerations}\
\putsssec{discuss_heuristics}{Generalizing \PNAME Heuristics}
We propose in this work heuristics for both schedule prioritization (SP) and resource partitioning (RP) for two concurrent kernels. Our SP heuristic can be extended to more kernels, that is, instead of two kernels, runtime can prioritize scheduling of multiple kernels again in the order of their resource requirements (number of workgroups), low to high. Similarly, our RP heuristic timing analysis can be extended to more kernels. As for two kernels, we can slowdown multiple kernels using the proposed analytical model and assess CU allocation that leads to best performance. However, this model does not factor in the increasing memory interference with more concurrent kernels and further investigation can be necessary in this regard. We leave evaluating these heuristics for more concurrent kernels to future work. 

\putsssec{caches}{Caching Considerations}
On MI300X, switching from CU-compute to DMA transfers does not incur any extra cache write-backs. This is so, as L2 is private per XCD (\figref{fig:bckg_mi300x}), it is already written-back to AMD Infinity Cache at GPU kernel completion. Further, sDMA engines read data from AMD Infinity Cache and write to HBM allowing subsequent GPU kernel to read sDMA transfer's output. That said, in alternate architectures, preceding GPU kernels can proactively write to common cache levels between compute and DMA engines preventing any cache-related overheads (e.g., cache write-back, flush) from happening on the critical path.

\putsssec{intra_node}{Inter-Node Communication Considerations}
Our study primarily addresses intra-node communication within an MI300X Infinity Platform node. While this is most prominent for inference and small-scale training, large-scale distributed ML training involves inter-node communication as well. In such cases, ML algorithms attempt to use both intra and inter node communication in tandem with attempting to maximize former as much as possible for better performance. This is done as intra-node bandwidth is significantly higher than inter-node bandwidth. In such cases, even with a focus on intra-node communication only, \OPNAME can be useful end-to-end. %\added{
Further, large-scale multi-node ML employ hierarchical collectives which break down collectives into intra and inter-node steps\cite{rajbhandari2022deepspeedmoeadvancingmixtureofexpertsinference} and ConCCL can be utilized for intra-node steps.
%} 

\putsssec{glocal}{Global/Local Optimization}
Our proposed heuristics/techniques can in some sense be classified as local optimization strategies. That is, we identify given two or more concurrent kernels, which of schedule prioritization or resource partitioning gets better performance. We understand that for better end-to-end performance for an application, with many kernels, such local decisions have to be combined at global-level to get better performance. Prior works in this regard~\cite{KatoLakshmanan2011-timeGraph} can be used in tandem with our work to do so.

\putsssec{power}{Power Considerations}
As GPUs get more capable, power constraints are increasingly at the forefront in their design. A power-agnostic scheduler could, by over-employing \PNAME, lower performance by causing GPU power to be stressed leading to power management events. Alternately, a more power-aware scheduler can employ \PNAME more judiciously by prioritizing concurrency for complementary power kernels. We leave design of such heuristics to future work. 

\putsssec{applicability}{Implications of CPU orchestration}
As discussed in \ssecref{isolated_conccl}, DMAs are orchestrated using CPU, the launch/sync cost for transfers are not amortized at smaller sizes. A GPU control-path might help solve this problem and we leave investigating software/hardware optimizations to address this as part of future work.

\putsec{related}{Related Work}

 %\noindent
Given the prevalence of \PNAME in ML, optimized \PNAME is key to continued scaling of ML.
In our work, we methodically study \PNAME, analyze resulting interference and most importantly, without making any invasive changes in existing system software and hardware, we identify techniques to use available software and hardware resources in optimized way to maximize benefit of \PNAME. 

\noindent
\textbf{Computation-Communication Kernel Overlap:}
Several GPU communication kernel libraries have been developed which improve collective performance in isolation - RCCL~\cite{rccl}, MSCCLang~\cite{cowan2023mscclang}, and more~\cite{nccl,shah2023taccl,Cheng_MSCCL_A_GPU-driven}. However, none of these take any special measures to mitigate any interference where collective kernels execute concurrently with computation kernels which are abundant in ML executions. While some prior works study this overlap algorithmically, they do not provide insights into the resource contention therein~\cite{patiToTC23}. Other works study \PNAME scenarios prevalent in DLRM recommendation~\cite{naumov2019deep} and Megatron-LM Transformer~\cite{ShoeybiPatwary2019-megatronlm} model executions using both microbenchmarks as well as end-to-end models executions~\cite{rashidi2021enabling}. Specifically, they study all-reduce slowdowns in concurrence with GEMMs, embedding table lookup and other backprop compute operations. Their study finds that all-reduce can slowdown on an average 1.4$\times$ in the Transformer model when overlapped with backprop operations, and by up to 6.2$\times$ when overlapped with both GEMMs and memory-bound embedding lookup in DLRM. Our work expands this to systematically study execution of collectives (from RCCL~\cite{rccl} which incorporates optimizations from other libraries for an AMD MI300X GPU) in concurrence with compute operations. It studies additional collectives such as all-gather and all-to-all, and furthermore, proposes a taxonomy to incorporate a wide-range of possible \PNAME scenarios. This taxonomy enables the study of \PNAME scenarios with different overlap characteristics (e.g., G-long, C-long) and includes communication-computation pairs that are not only prevalent in models today, but can be possible in future models and/or distributed setups. Finally,  we not only study these contentions, but also develop two optimizations including, schedule prioritization and compute (CU) partitioning, that improve end-to-end performance of \PNAME without any additional accelerator and/or hardware changes. While there have been numerous works on improving concurrent kernel execution on GPUs via appropriate resource partitioning~\cite{adriaens2012case,jiao2015improving}, prioritization~\cite{KatoLakshmanan2011-timeGraph} and concurrency-aware kernels~\cite{pai2013improving,MaXie2020-rammer,pati2024global}, they primarily focus on multiple compute kernels. Our work applies these techniques to coarse-grain compute-communication kernels. NanoFlow~\cite{zhu2024nanoflow} also performs resource partitioning for \PNAME; however, they create very fine-grained kernels by slicing ML input from micro into nano-batches, and find necessary resource allocation from a huge and complex search space unlike our simple yet useful lookup based heuristic. 

\noindent
\textbf{Communication Offload:}
Works which improve \PNAME performance usually offload communication to dedicated customized accelerators on GPUs. For example, ACE~\cite{rashidi2021enabling} is a custom accelerator for offloading communication which can free up compute units for concurrent compute. It further buffers intermediate, partially reduced data to avoid their reads and writes to memory which reduces memory interference. %It further buffers intermediate, partially reduced, data to avoid their reads and writes to memory in a ring-like allreduce algorithm. The latter reduces memory interference, however only benefits allreduce collective. 
Similarly, other works  
%SP: technically, they can help with AG too - sDMAs have to read the same data muliple times to broadcast to different devices. Can buffers in ACE enable it to do it in one go with a direct mechanism? Or can buffers enable them to only write and pass it along in ring, avoiding reads? Not discussed in paper, wonder if we need to bring it up in discussions. 
require extensive hardware support such as compute-capable switches to offload reduction operations while reducing volume of data moved over network links and memory sub-system for all-reduce~\cite{klenk2020network}. 
Furthermore, switch offloads still require GPU kernels to orchestrate data movement and in-switch commands, leading to interference with concurrent computation. We consider offload but by leveraging \textit{existing} data-movers (DMAs), requiring no hardware changes and show \PNAME benefits with two key ML collectives, all-gather and all-to-all. 

\noindent
\textbf{Communication Offload to DMAs:}
There are also work which offload communication to DMA engines. For example, MSCCL++~\cite{Cheng_MSCCL_A_GPU-driven} offloads larger size collective operations to DMA engines, however it initiates DMA call from GPU involving CU (through proxy channel in CPU), which in our work we prefer to reserve only for compute operations. ARK~\cite{hwang2023ark} suggests communication offload to DMA but emphasizes on DMA call overhead time and involves GPU threads to reduce such overhead either by significant SW changes or by addition of new DMA prototype; we, in contrast rely on existing software stack and HW and still show benefit of sDMA offload. Some of the works~\cite{async-tp} leverage DMAs for fine-grained \PNAME. However, unlike this work, they provide limited insights into challenges with such overlap and therefore provide limited insights into the potential for DMAs alone to help overcome \PNAME interference like this work highlights with ConCCL. Other works require hardware modifications to initiate DMA transfers in a fine-grained manner~\cite{pati2024t3}.

\noindent
\textbf{Fine-grained Computation-Communication Overlap:}
Other interesting works discuss techniques for fine-grain overlap of compute/communication to improve concurrency~\cite{punniyamurthy2023optimizing, jangda2022breaking, wang2022overlap,zhu2024nanoflow,pati2024t3,async-tp}. While such fine-graining techniques are promising, they necessitate pervasive changes in SW/HW to realize overall performance benefit. For example, Fused Computation-Collective~\cite{punniyamurthy2023optimizing} fuses communication operation directly in GEMM kernel requiring exponential implementations in GEMM library.  
\putsec{conclusion}{Conclusion}
In this work, we carefully characterize the performance of concurrent computation and communication (\PNAME), an important ML paradigm, on state-of-art MI300X GPU and observe that speedups attained are, not unexpectedly, on average only 21\% 
of ideal speedup possible due to interference between concurrent GPU kernels. To improve performance of this important primitive, we first evaluate dual strategies of schedule prioritization and careful resource partitioning of compute units on GPUs to push performance attained with \PNAME (on average 42\% of ideal speedup). Additionally, we provide heuristics that can guide a GPU runtime to harness these strategies. To further improve speedups for \PNAME, we offload communication to DMA engines via building \OPNAMEBOLD (Concurrent Communication Collectives) proof-of-concepts. We demonstrate how \OPNAME  closes the gap between realized and ideal speedup for \PNAME (on average about 72\%  of ideal speedup is realized, up to 1.67$\times$ speedup). We believe DMA engines have an important role to play for \PNAME performance and argue for their continued betterment. 

\section*{Acknowledgment}
We thank our colleagues Wenkai Du, Gilbert Lee, Alexander Kaganov, Leo Dong, Anthony Asaro, Padmini Nujetti for many helpful discussions. We also thank Nicholas Malaya, Alex Habeger, Kevin Cheng, Gowri Shankar Guttikonda for help with equipment. Finally, we also thank Gabriel Loh and the anonymous ISPASS reviewers for helping improve the paper. AMD, the AMD Arrow logo, AMD ROCm, AMD Instinct, AMD Infinity Cache, AMD Infinity Fabric, and combinations thereof are trademarks of Advanced Micro Devices, Inc. Other product names used in this publication are for identification purposes only and may be trademarks of their respective companies.

%%%%%%% -- PAPER CONTENT ENDS -- %%%%%%%%

% \balance
\bibliographystyle{IEEEtran}
\bibliography{ref}

\end{document}